\documentclass[%
 reprint,
 amsmath,amssymb,
 aps,
prl,
superscriptaddress,
]{revtex4-2}

\usepackage{graphicx}
\usepackage{dcolumn}
\usepackage{bm}
\usepackage[utf8]{inputenc}
\usepackage{amsmath}

\usepackage{derivative}
\usepackage{braket}
\usepackage{xcolor}
\usepackage{microtype}
\usepackage{siunitx}
\usepackage{booktabs}
\usepackage{hyperref}

\begin{document}

\preprint{APS/123-QED}

\title{Entangling ions with engineered light gradients}
\author{Tommaso Faorlin}
\affiliation{Universität Innsbruck, Institut für Experimentalphysik, Technikerstraße 25/4, 6020 Innsbruck, Austria}
\email{tommaso.faorlin@uibk.ac.at}

\author{Lorenz Panzl}%
\affiliation{Universität Innsbruck, Institut für Experimentalphysik, Technikerstraße 25/4, 6020 Innsbruck, Austria}

\author{Phoebe Grosser}%
\affiliation{Universität Innsbruck, Institut für Experimentalphysik, Technikerstraße 25/4, 6020 Innsbruck, Austria}

\author{Pablo Viñas}
\affiliation{Instituto de Física Teórica (UAM-CSIC), Universidad Autónoma de Madrid, Cantoblanco, 28094 Madrid, Spain}

\author{Alan Kahan}
\affiliation{Instituto de Física Teórica (UAM-CSIC), Universidad Autónoma de Madrid, Cantoblanco, 28094 Madrid, Spain}

\author{Walter Joseph Hörmann}%
\affiliation{Universität Innsbruck, Institut für Experimentalphysik, Technikerstraße 25/4, 6020 Innsbruck, Austria}

\author{Yannick Weiser}%
\affiliation{Universität Innsbruck, Institut für Experimentalphysik, Technikerstraße 25/4, 6020 Innsbruck, Austria}

\author{Giovanni Cerchiari}%
\affiliation{Universität Innsbruck, Institut für Experimentalphysik, Technikerstraße 25/4, 6020 Innsbruck, Austria}
\affiliation{Naturwissenschaftlich-Technische Fakultät, Universität Siegen, Walter-Flex-Straße 3, 57068 Siegen, Germany}

\author{Thomas Feldker}
\affiliation{Alpine Quantum Technologies, Technikerstraße 17, 6020 Innsbruck, Austria}

\author{Alexander Erhard}
\affiliation{Alpine Quantum Technologies, Technikerstraße 17, 6020 Innsbruck, Austria}

\author{Georg Jacob}
\affiliation{Alpine Quantum Technologies, Technikerstraße 17, 6020 Innsbruck, Austria}

\author{Juris Ulmanis}
\affiliation{Alpine Quantum Technologies, Technikerstraße 17, 6020 Innsbruck, Austria}

\author{Rainer Blatt}%
\affiliation{Universität Innsbruck, Institut für Experimentalphysik, Technikerstraße 25/4, 6020 Innsbruck, Austria}
\affiliation{Alpine Quantum Technologies, Technikerstraße 17, 6020 Innsbruck, Austria}
\affiliation{Institute for Quantum Optics and Quantum Information of the Austrian Academy of Sciences, Technikerstraße 21a, 6020 Innsbruck, Austria}

\author{Alejandro Bermudez}%
\affiliation{Instituto de Física Teórica (UAM-CSIC), Universidad Autónoma de Madrid, Cantoblanco, 28094 Madrid, Spain}

\author{Thomas Monz}
\affiliation{Universität Innsbruck, Institut für Experimentalphysik, Technikerstraße 25/4, 6020 Innsbruck, Austria}
\affiliation{Alpine Quantum Technologies, Technikerstraße 17, 6020 Innsbruck, Austria}

\date{\today}

\begin{abstract}
Spectral crowding of collective motional modes limits the fidelity of entangling interactions in trapped-ion quantum processors by inducing off-resonant coupling to spectator modes. We introduce a geometric-phase entangling interaction driven by a transverse, time-dependent structured-light force. By applying the force in a plane orthogonal to the optical propagation direction, we reduce the effects of spectral crowding while preserving single-ion addressing. The scheme is compatible with arbitrary qubit encodings, provided that the qubit states experience a differential AC Stark shift. We experimentally realise high-fidelity two-qubit gates with error rates below $\SI{5e-3}{}$ in ion crystals containing up to 12 ions confined within a single potential well. These results establish gradient-field light-shift gates as a scalable approach to high-fidelity entanglement generation in spectrally crowded trapped-ion systems.
\end{abstract}

\maketitle

\begin{figure*}[t!]
\centering
\includegraphics[width=\textwidth]{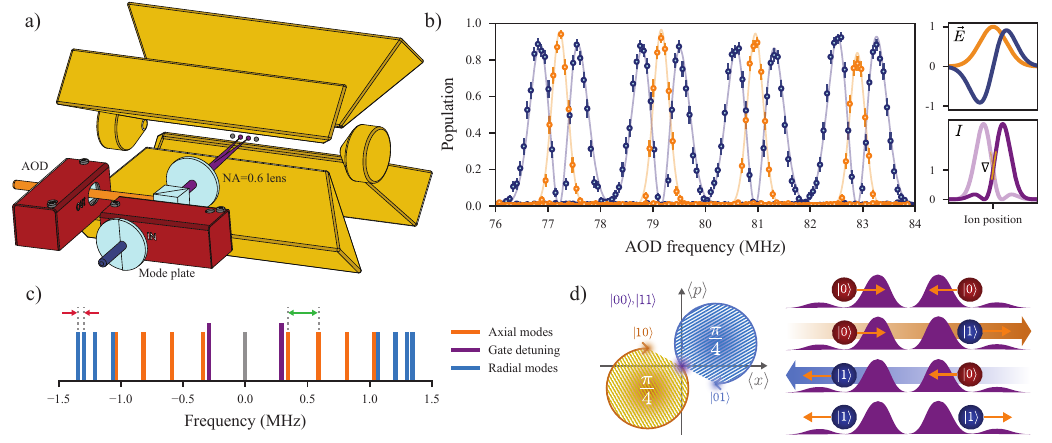}
\caption{a) Sketch of the addressing system. Two beams in two spatial modes (orange and blue) are overlapped on a polarizing beam splitter cube, after being respectively deflected by AODs. The blue beam, initially propagating like a Gaussian beam, is converted to a TEM$_{10}$ spatial mode with a phase plate before entering the AOD. An NA=0.6 lens focuses the two co-propagating light fields at the ions' position. b) Deflector scan. The frequency of each AOD is swept over the ion chain. Each point is the result of a Ramsey experiment, since the off-resonant light at \SI{532}{\nano\metre} is applied between two $\pi/2$ pulses before measuring the population. The orange peaks decrease in height from left to right due to an imperfect alignment of the laser beam. The right upper plot shows the electric fields for the two transverse motional modes. The bottom plot instead, displays the resulting intensity profile at the ions due to their interference. Depending on the phase between the two fields, the intensity peak can be on the right or on the left of the ion position, generating an alternating gradient. c) Motional spectrum around a carrier transition corresponding to measured values. The spacing between the two highest radial modes depends on the ratio between the two trapping frequencies, while for the lowest two axial modes the ratio is fixed to $\sqrt{3}$. d) Phase space picture and geometric phases. If two ions are illuminated with the two beams oscillating in phase, the $\ket{01}$ and $\ket{10}$ quantum states will be displaced and perform loops in phase space, acquiring a geometric phase, while the force on $\ket{00}$ and $\ket{11}$ cancels out.}
\label{fig:panel}
\end{figure*}

\emph{Introduction.---} 
The ability to selectively drive well-defined transitions while suppressing unwanted couplings is a central challenge in the coherent control of quantum systems. As quantum processors scale in size and complexity, spectral crowding and parasitic interactions increasingly limit achievable gate fidelities. This issue arises across multiple platforms, including silicon quantum dots~\cite{Zajac2018}, nitrogen-vacancy centres in diamond~\cite{Bradley2019} and superconducting transmon processors~\cite{Malekakhlagh2020, Cai2021}, where off-resonant excitation of unwanted transitions can degrade performance. Dynamical decoupling (DD) techniques~\cite{Rahman2024, Hahn1950} provide a powerful method to mitigate such effects~\cite{Kofler2025, Cheng2023, Milne2020, Kang2021, Ezzell2023}, but typically at the expense of increased control complexity and experimental overhead.
Trapped-ion systems offer a leading architecture for quantum information processing due to their long coherence times and high-fidelity operations. In these systems, ions are confined by static and radio-frequency electric fields, forming linear crystals whose collective motion can be described in terms of quantised normal modes. Selected motional modes serve as a quantum data bus that mediates entangling interactions between qubits. However, as the number of ions increases, the frequency separation between adjacent motional modes decreases. Entangling gates that rely on spectrally addressing one or a small subset of these modes therefore suffer from parasitic coupling to nearby spectator modes, limiting coherent control and scalability.\\~
Here, we present an alternative approach to suppressing off-resonant motional couplings by engineering the spatial structure of the driving field rather than relying on dynamical decoupling. We implement a light-shift (LS) entangling gate based on electric field gradients, generated by a tightly focused laser beam in a TEM$_{10}$-like transverse mode (Fig.~\ref{fig:panel}-a,b). By interfering optically structured beams, we create a state-dependent optical dipole force aligned with the axis of a linear ion crystal, while the beam propagates transversely to it. This configuration enables targeted addressing of selected ion pairs with minimal crosstalk and mediates a conditional geometric phase interaction through the axial motional spectrum, operating in a spectral region that is less susceptible to coupling to non-targeted modes. Previous high-fidelity demonstrations of LS gates were restricted to two-ion chains under collective illumination~\cite{Clark2021, Baldwin2021, Leibfried2003}. In contrast, our approach preserves single-ion addressing while extending the interaction to longer chains~\cite{Kim2008, Baldwin2021, Sawyer2021}. This capability is particularly relevant in light of architectural optimisation studies that identify ion chains of approximately 12 ions as a favorable operating point, balancing connectivity, shuttling overhead, and control complexity in surface trap architectures~\cite{Murali2022}. We experimentally demonstrate high-fidelity entangling operations with fidelities above 99.5\% in ion chains with up to 12 ions, establishing gradient-field LS gates as a scalable approach for trapped-ion quantum processors.
\par
\smallskip
\emph{Entangling interaction.---} In trapped-ion systems, three sets of quantised motional modes arise along the three spatial axes. The mode spacing, defined as the frequency separation between adjacent normal modes, depends on the direction. In one-dimensional, harmonically-confined ion chains, axial modes are more widely spaced than radial modes~\cite{James1998}.
Furthermore, the spacing of axial modes does not depend on the ion number, whereas the mode spectrum becomes increasingly crowded in the radial direction. This, in combination with the larger mode spacing, makes axial modes particularly suitable for mediating entanglement between ion pairs because the coupling to parasitic motional modes is reduced. An entangling gate duration of $\mathrm{t}_{\mathrm{gate}} = \SI{100}{\micro\second}$ requires a detuning of $\delta = 2\pi\times\SI{10}{\kilo\hertz}$ from a motional mode. We consider a typical trap configuration with axial and radial centre-of-mass (COM) frequencies of $\omega_\text{COM}^{\text{(ax)}} =2\pi\times\SI{341.7}{\kilo\hertz}$ and $\omega_\text{COM}^{\text{(rad)}} \approx2\pi\times\SI{1.33}{\mega\hertz}$, respectively. When targeting the COM mode, the relative separation to the nearest neighbouring mode is $70\,\delta$ in the axial direction, compared to $3\,\delta$ radially (Fig.~\ref{fig:panel}-c). We refer to the nearest axial neighbouring mode as \textit{breathing} (BR) mode.\\~
We leverage axial modes for entanglement while preserving single-ion addressing by employing transverse vector forces produced by two off-resonant interfering laser beams with optically structured spatial profiles, corresponding to the TEM$_{00}$ and TEM$_{10}$ modes~\cite{Erhard2026}. With a relative frequency difference, the two beams generate an oscillating transverse AC Stark shift gradient at the ions' position (Fig.~\ref{fig:panel}-b).  If the beams exhibit $\text{lin}\parallel\text{lin}$ polarisation, the AC Stark shift gradient is induced by an intensity gradient and is differential for states separated by an optical transition. Conversely, if the beams exhibit $\text{lin}\perp\text{lin}$ polarisation, the AC Stark shift gradient is induced by a polarisation gradient and is differential for states with different orbital angular momentum. The gradient induces an optical dipole force (ODF), which can be used to generate entanglement via the accumulation of state-dependent geometric phases.
Our approach is intrinsically compatible with \textit{omg}-qubit (optical, metastable, ground-state qubits) implementations~\cite{Allcock2021}. The only requirement on the electronic states chosen to encode the qubit is the presence of a differential AC Stark shift, rendering them sensitive to the oscillating transverse gradient. We note that Mølmer–Sørensen–type (MS) transversal entangling interactions have been proposed and realized on hyperfine and optical qubits, using different beam architectures~\cite{Cui2025, Mazzanti2023, West2021, Mai2025}.
We remark that related work demonstrated the excitation of axial motional modes in a single-ion system using vortex beams carrying optical angular momentum~\cite{Stopp2022}.\\~
Intrinsically, the two-qubit gate implements a $\sigma_z \otimes \sigma_z$ interaction, which can be generalised to an arbitrary $\sigma_\phi=\sigma_x\cos{\phi}+\sigma_y\sin{\phi}$ interaction with local single-qubit rotations~\cite{Foss-Feig2025}. We drive the motion of the ions close to one of the motional mode frequencies $\Delta=\omega_\text{COM}^{\text{(ax)}}-\delta$, forcing the two-qubit states of the addressed ion pair to undergo state-dependent motional trajectories in position-momentum space (Fig.~\ref{fig:panel}-d). If a two-qubit state $i\in(00,01,10,11)$ experiences a net ODF $F_i$, the modulation results in the state being displaced along a circular path of radius $R\propto|F_i|$ and periodicity $2\pi/\delta$. At time $t=2\pi/\delta$ the two-qubit system will return to its original motional state with an acquired geometric phase of $\Phi_i \propto 2\pi|F_i|^2$. The phase-space trajectories coincide with those of a classical forced harmonic oscillator, while the geometric phase corresponds to the Berry phase~\cite{Ozer2011}. Ideally (perfect loop closure), for a judicious choice of ODF strength, geometric phases of $\Phi_{00} = \Phi_{11}$ and $\Phi_{01} = \Phi_{10} = \Phi_{00} \pm \frac{\pi}{2}$ can be acquired, which corresponds to the ideal fully-entangling two-qubit gate. The presence of additional motional modes, with their respective frequency difference from $\Delta$, causes additional state-dependent motional trajectories, resulting in imperfect loop closures~\cite{Kofler2025}. Note that the total phase needed for entanglement can be acquired by performing multiple loops in phase-space with a smaller displacement from the origin~\cite{MolmerSorensen1999}.\par
\smallskip
\emph{Experimental configuration and results.---}
The setup is similar to~\cite{Pogorelov2021}, optimised for Barium ions. A chain of $^{138}\mathrm{Ba}^+$ ions is loaded via laser ablation into a linear Paul trap with secular trap frequencies $(\omega^{(x)},\, \omega^{(y)},\, \omega^{(z)})=2\pi\times(1.304,\, 1.344,\, 0.3422)\,\mathrm{MHz}$. A magnetic field of approximately $4.5\,\mathrm{G}$, generated by permanent magnets, is orientated perpendicular to the ion chain to induce a ground state (GS) ($6\mathrm{S}_{1/2}$) Zeeman splitting of $\SI{12.49}{\mega\hertz}$. Collective radio-frequency (RF) control of the GS is implemented with an antenna impedance-matched to the frequency of this splitting. For the addressed ODF-beams, far off-resonant light  (\SI{532}{\nano\metre}) is applied (Coherent Verdi V5). The light is split into two channels and fed into separate AOMs in a double-pass configuration to control the detuning between the beams, before being sent towards the addressing units (Fig.~\ref{fig:panel}-a). In one beam path, a phase element~\footnote{Holoor Hermite-Gauss mode converter PE202-Q-Y-A} introduces a $\pi$-phase shift between two halves of the incident beam, converting the TEM$_{00}$ to a TEM$_{10}$ mode at the ion's position. 
The beams are routed through two independent acousto-optical deflectors (AODs), each capable of simultaneously addressing two ions and independently controlling the phase of each ODF beam. Since AODs operate most efficiently with linearly polarised light, this configuration enables operation with either parallel or orthogonal linear polarizations for the two beams, inducing an intensity or a polarisation gradient. The beams are then overlapped and expanded before reaching an objective lens with a numerical aperture (NA) of $0.6$, situated in front of the trap.\\~ 
\begin{figure}[t!]
\centering
\includegraphics[width=.48\textwidth]{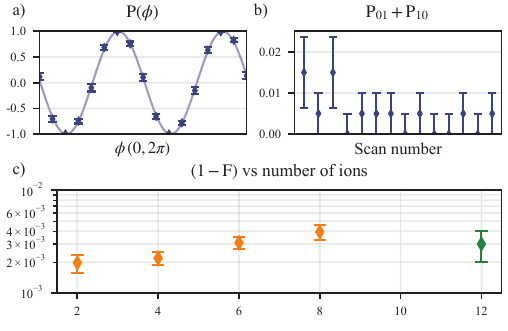}
    \caption{a) Parity oscillations for a single gate on a two-ion crystal. Error bars are given by $\sqrt{(1-\mathrm{P}(\phi)^2)/\mathrm{N}_{\mathrm{shots}}}$. b) Residual spin motion entanglement. Each point is the result of $\mathrm{N}_{\mathrm{shots}}=200$ measurements of the Bell state at the closure of the gate, and error bars are given due to projection noise. The error bar on the points at 0 are defined via the rule of succession, i.e. $1/(\mathrm{N}_{\mathrm{shots}}+1)$. c) Gate infidelity vs chain length. Each point for $N>2$ is estimated by averaging the infidelities of gates performed on the innermost and outermost ion pairs. The fidelity in each configuration is extracted from an exponential decay fit to the fidelity of 1, 3, 5, 7 and 9 concatenated gates, in order to account for state-preparation and measurement (SPAM) errors. The orange data points are measured with $^{138}$Ba$^+$. The green point is measured by co-authors A.E. and T.F. on a $^{40}$Ca$^+$ based trapped ion system (two loops in phase space and $\mathrm{t}_{\mathrm{gate}}=\SI{120}{\micro\second}$).}
    \label{fig:results}
\end{figure} 
We encode qubits in the $\ket{0} = \ket{m_j=-1/2}$ and $\ket{1} = \ket{m_j=+1/2}$ Zeeman sublevels of the GS.
The ions are first cooled to the Doppler limit ($\bar{n}_\text{COM}^{\text{(ax)}} \approx 22$) via the $6\mathrm{S}_{1/2}\to 6 \mathrm{P}_{1/2}$ transition. The radial modes are cooled via electromagnetically-induced transparency (EIT) cooling on the $6\mathrm{S}_{1/2}\to 6 \mathrm{P}_{1/2}$ transition. The two lowest-order (COM and BR) axial modes are cooled further via resolved sideband cooling ($\bar{n}_{\text{COM}}^{\text{(ax)}} \approx 0.12$) on the $6\mathrm{S}_{1/2}\,(m_j=-1/2)\to 5 \mathrm{D}_{5/2}\,(m_j=-5/2)$ transition. The ions are initialised to the $\ket{0}$ state by two-step, frequency-selective optical pumping through the $5 D_{5/2}$ and $6 P_{3/2}$ manifolds. This process prepares all $N$ qubits in the $\bigotimes_N \ket{0}$ state, after which the two-qubit gate can be applied on the desired ion pair. Discrimination between the two qubit states is performed by shelving the $\ket{1}$ state to the $5 \mathrm{D}_{5/2}$ manifold via collective, narrow-linewidth pulses, driving a complete population transfer ($\pi$ pulse). Finally, the quantum state is detected via fluorescence scattering on the short-lived $6 \mathrm{S}_{1/2} \rightarrow 6\mathrm{P}_{1/2}$ transition. The relative detuning $\Delta$ of the ODF beams is set via the AOMs, a respective phase difference is controlled with the AODs, such that a geometric phase is accumulated in the $\ket{01}$ and $\ket{10}$ states. Because the two-qubit gate induces a $\sigma_z \otimes \sigma_z$ interaction, we map phases acquired by the two states into population imbalances via a basis change~\cite{Clark2021, Leibfried2003} (Ramsey-like sequence), including a Hahn~\cite{Hahn1950} echo to symmetrise the spin-dependent phase acquired through the driven evolution of the $\ket{0}$ and $\ket{1}$ states~\cite{Ballance2016}, as well as to compensate for single-qubit Stark shifts. In our measurements, a gate consists of 4 phase-space loops at $\delta=2\pi\times\SI{20}{\kilo\hertz}$, with a $\mathrm{t}_\text{gate} = 4 \cdot (2 \pi / \delta)\approx\SI{220}{\micro\second}$. The deviation from the ideal gate time $\mathrm{t}_\text{gate}=\SI{200}{\micro\second}$ is due to pulse shaping. The results are reported in Fig.~\ref{fig:results} in terms of parity oscillations and residual spin-motion entanglement for a gate on a two-ion crystal, and the infidelity versus the number of ions in the trap.\\~
We verify the entangled nature of the output state via observation of parity oscillations (Fig.~\ref{fig:results}-a) $\mathrm{P}(\phi)=\mathrm{P}_{00}(\phi)+\mathrm{P}_{11}(\phi)-\mathrm{P}_{01}(\phi)-\mathrm{P}_{10}(\phi)\propto \lvert \mathrm{A}\lvert\sin{\phi}$, obtained by varying the analysis phase $\phi$ of an additional $\pi/2$ pulse~\cite{Leibfried2003}. We extract the amplitude $\lvert \mathrm{A}\lvert$ via a sine fit. We define the state fidelity by measuring the overlap $\mathrm{F}=\braket{\psi\lvert\rho\lvert\psi}=(\mathrm{P}_{00}+\mathrm{P}_{11})/2+\lvert \mathrm{A}\lvert/2$ with the ideal Bell state (Fig.~\ref{fig:results}-b) $\psi=(\ket{00}-i\ket{11})/\sqrt{2}$ at $\mathrm{t}_{\mathrm{gate}}$. In order to distinguish the fidelity of the gate operations from errors introduced via SPAM, we apply multiple gate operations in succession and observe exponentially decaying fidelities~\cite{Ballance2016, Hrmo2023}. The average fidelity (and the corresponding average gate error) is then given by the decay constant of this exponential.
In Fig.~\ref{fig:results}-c, we plot the gate infidelity as a function of the ion-chain length. For each chain with $N>2$, the value is obtained by averaging the results for the innermost and outermost ion pairs. Errors due to SPAM are excluded, corresponding to $0.2(2)$\%. Our measurements show that light-based coherent entanglement generation with fidelities exceeding $99.5\%$ is achievable even in medium-sized ion crystals. The observed infidelities lie below the fault-tolerance threshold, enabling the encoding of logical qubits and the implementation of quantum error correction (QEC) protocols~\cite{Barends2014, Ye2025}.

\emph{Error budget.---}
Table~\ref{tab:error_budget} summarises the dominant error sources limiting the fidelity of a $\mathrm{t}_\mathrm{gate}=\SI{220}{\micro\second}$ gate performed on the GS of a two-ion crystal. The infidelity due to parasitic coupling to axial spectator modes is in the order of $10^{-4}$. In contrast, the same gate performed on radial modes would yield an infidelity of $7.4\times10^{-2}$ (increasing with ion number), highlighting the suppression of spectator-mode coupling as the principal advantage of the transverse gradient gate. The first four error contributions (non-unitary/incoherent errors) were determined by numerical simulation of the Lindblad master equation for a density matrix $\rho$,
\begin{equation}
    \dot{\rho} = - \frac{i}{\hbar} [H_I, \rho] + \sum_i \gamma_i \left( L_i \rho L_i^\dagger - \frac{1}{2} \{ L_i^\dagger L_i, \rho \} \right) \, ,
\end{equation}
where $\gamma_i$ are the damping rates, $L_i$ the jump operators and $H_I$ the Hamiltonian for the forced quantum harmonic oscillator (in the interaction picture),
\begin{equation}
    H_I = - \frac{i}{2} \sum_{ \substack{i \in \{00, 01, \\ 10, 11 \} }} F_i \left( a e^{-i \delta t} + a^\dagger e^{i \delta t} \right) \ket{i} \bra{i} \, .
\end{equation}
In the last equation, $a$ and $a^\dagger$ are the annihilation and creation operators, respectively. Non-unitary errors were characterised by including the corresponding Lindbladian jump operator $L_i$ in the master equation simulation with the corresponding rate~\footnote{The contribution of these incoherent error sources  is estimated separately and added to the error budget linearly, which assumes that we are in the high-fidelity regime.}. In the simulation, we consider ground-state lifetime ($T_1$) of $\SI{7}{\second}$, ground-state coherence time ($T_2$) of $\SI{200}{\milli\second}$, COM motional coherence time of $\SI{40}{\milli\second}$, and COM heating rate of 1.5 phonons/ion/s. The coherence times were determined via Ramsey interferometry. Note that the motional coherence time was measured for an eight-ion crystal. The contribution to gate infidelity due to motional decoherence becomes prominent with the ion chain growing in length, dominated by motional heating, which scales linearly with the ion number~\cite{Brownnut2015}. The next axial motional mode ($\omega_\mathrm{BR}^{\mathrm{(ax)}}$) exhibits a reduced heating rate; however, due to inhomogeneous participation in the collective motion, it is not suitable for arbitrary ion pairs.
All other error contributions were experimentally characterised. The sensitivity of the gate to rotation errors in the collective RF pulses (performed at the beginning, middle, and end of the gate) was characterised via randomised benchmarking of the Clifford single-qubit unitaries. The off-resonant scattering error, the fundamental limit of this gate, is split into the contribution from elastic (Rayleigh) and inelastic (Raman) scattering~\cite{Ozeri2007, Moore2023}. This is dominated by the TEM$_{00}$ mode, which has a non-zero intensity at the ion's position. We currently measure an error contribution at the $10^{-4}$ level, measured by applying the gate sequence without performing the RF pulses and checking how much of the initial population ends up in the $\ket{1}$ and the D manifold levels. As the force on the ions is directly dependent on the gradient, one can achieve the same ODF with a different choice of TEM$_{10}$ mode optical power (compensating with the TEM$_{00}$ mode optical power) and thereby lower the total scattering rate even further.
\begin{table}[t]
\centering
\caption{Error budget for $\mathrm{t}_\mathrm{gate}=\SI{220}{\micro\second}$ implemented with four phase-space loops ($\delta = 2\pi \times \SI{20}{\kilo\hertz}$) on GS qubits in an eight-ion crystal. The same gate is evaluated using an axial or a radial motional mode. The dominant error source, i.e. the one caused by unwanted excitation of spectator modes, is suppressed in the axial implementation.}
\label{tab:error_budget}
\vspace{0.2cm}
\begin{tabular*}{\columnwidth}{@{\extracolsep{\fill}} l S[table-format=3.1] S[table-format=5.1]}\hline\hline
\textbf{Error mechanism}&
\multicolumn{2}{c}{\textbf{Infidelity} ($\times 10^{-4}$)} \\
& \multicolumn{1}{c}{Axial} & \multicolumn{1}{c}{Radial} \\
\hline
\hline
Spectator modes            & 2.1 & 742.4 \\
Qubit $T_2$ decoherence    & 9.6 & 9.6 \\
Motional decoherence       & 8.7 & 18.8 \\
RF pulses                  & 2.4 & 2.4 \\
COM mode heating           & 4.0 & 0.9 \\
Qubit $T_1$ decay          & 1.1 & 1.1 \\
Scattering (Rayleigh \& Raman)           & 1.2 & 1.2 \\
\hline
\textbf{Total}              & 29.1 & 776.4 \\
\hline\hline
\end{tabular*}
\end{table}

\emph{Conclusions and discussion.---}
We demonstrated an LS–mediated entangling interaction with error rates at the $10^{-3}$ level in crystals of up to 12 ions, achieving the threshold requirements for fault-tolerant QEC~\cite{Barends2014, Ye2025}. Future improvements may include implementing different DD sequences, testing different numbers of phase-space loops, and transitioning to hyperfine encoding in $^{137}\mathrm{Ba}^+$, to suppress error contributions arising from finite qubit coherence time ($T_2$). We envision further improvements to the experimental apparatus to enhance motional coherence and GS lifetime. Furthermore, we plan to increase the radial confinement in order to achieve longer ion strings without further lowering the axial confinement. Beyond quantum computing, coherent control of force vectors transverse to the propagation of light enables single- or multi-ion entangling interactions also in two-dimensional ion geometries. The static phase plate may be replaced by a spatial light modulator, allowing programmable control of the force-gradient orientation and selective excitation of motional modes along arbitrary directions. Furthermore, high-fidelity entanglement performed on a trapped ion's GS can also benefit quantum network nodes, as this specific encoding is also chosen for long-lived quantum information memories. Finally, we remark that the required optics could be directly integrated into a trap chip~\cite{Badawi2025}.

\emph{Acknowledgments.---}
We gratefully acknowledge support by the European Union’s Horizon Europe research and innovation program under Grant Agreement Number 101114305 (“MILLENION-SGA1” EU Project), by the Austrian Science Fund (FWF Grant-DOI 10.55776/F71) (SFB BeyondC) as well as the Austrian Research Promotion Agency (FFG) under contracts No. 897481 (High-Performance integrated Quantum Computing) and No. 896213 (Ion trap arrays for quantum computing) and by IQI GmbH. We further acknowledge funding by the Intelligence Advanced Research Projects Activity (IARPA) and the Army Research Office, under the Entangled Logical Qubits programme through Cooperative Agreement Number W911NF-23-2-0216. P.V., A.K. and A.B. acknowledge support from PID2021-127726NB-I00 and PID2024-161474NB-I00 (MCIU/AEI/FEDER,UE), from QUITEMAD-CM TEC-2024/COM-84, from the Grant IFT Centro de Excelencia Severo Ochoa CEX2020-001007-S funded by
MCIN/AEI/10.13039/501100011033, and from the CSIC Research Platform on Quantum Technologies PTI-001. This work is partially funded by the European Commission–NextGenerationEU, through Momentum CSIC Programme: Develop Your Digital Talent. A.E., T.F. and J.U. acknowledge support from the AWS (Austrian Promotional Bank) with funds from the National Foundation for Research, Technology and Development (Fonds Zukunft Österreich). G.C. acknowledges the support of the Ministry of Culture and Science of North Rhine-Westphalia. G.C. and Y.W. acknowledge the support of the Austrian Science Fund (FWF) (Crossref Funder ID 501100002428) project SONATINA (Grant DOI 10.55776/P36233).

The views and conclusions contained in this document
are those of the authors and should not be interpreted as
representing the official policies, either expressed or implied, of IARPA, the Army Research Office, or the U.S.
Government. The U.S. Government is authorized to reproduce
and distribute reprints for Government purposes
notwithstanding any copyright notation herein.


\bibliographystyle{apsrev4-2}
\bibliography{bibliography}

\end{document}